\documentclass[superscriptaddress,twocolumn,preprintnumbers,amsmath,amssymb]{revtex4}

\usepackage{graphicx}
\usepackage{dcolumn}
\usepackage{epsfig}
\begin{document}

\title{Flow behavior of colloidal rod-like viruses in the nematic phase}

\author{M. Paul Lettinga}
\affiliation{%
IFF, Institut Weiche Materie, Forschungszentrum J\"{u}lich,D-52425
J\"{u}lich, Germany }

\author{Zvonimir Dogic} \affiliation{Rowland Institute at Harvard, Harvard University, Cambridge MA 02142, USA}%

\author{Hao Wang} \affiliation{Department of Physics, North Dakota State University,
Fargo, ND 58105, USA}
\author{Jan Vermant}
\affiliation{Department of Chemical Engineering, Katholieke Universiteit Leuven,
 de Croylaan 46, B-3001 Leuven, Belgium\\
}%

\begin{abstract}
The behavior of a colloidal suspension of rod-like  {\it fd}
viruses in the nematic phase, subjected to steady state and
transient shear flows is studied. The monodisperse nature of these
rods combined with relatively small textural contribution to the
overall stress make this a suitable model system to investigate
the effects of flow on the non-equilibrium phase diagram.
Transient rheological experiments are used to determine the
critical shear rates at which director tumbling, wagging and
flow-aligning occurs. The present model system enables us to study
the effect of rod concentration on these transitions. The results
are in quantitatively agreement with the Doi-Edwards-Hess model.
Moreover, we observe that there is a strong connection between the
dynamic transitions and structure formation, which is not
incorporated in theory.
\end{abstract}

\maketitle

\section{Introduction}

When subjected to shear flow liquid crystals can exhibit a variety
of surprising phenomena, which arise because of the anisotropic
shape of the constituent rods. Theoretically the behavior of a
suspension of hard rods during shear flow can be described by the
equation that governs the time development of their probability
distribution function, as derived by Hess~\cite{Hess76} and by Doi
and Edwards~\cite{Doi86}. In the absence of a flow, the
Doi-Edwards-Hess (DEH) theory reduces to the Onsager description
of equilibrium nematic liquid crystals and can be used to describe
the isotropic-nematic (I-N) phase transition of a hard rod
suspension~\cite{Onsager49}. The rheological properties are
predicted to be highly non-linear functions of the P\'{e}clet
number (Pe), which is the ratio of shear rate $\dot{\gamma}$ over
rotational diffusion constant $D_{r}$. This is not surprising as
the Pe number can be much larger than unity when the rod-like
molecules have large aspect ratios.

The nonlinear response of the rheological properties indicates
that the shear flow distorts the equilibrium distribution of
macromolecules or rods. The spatiotemporal microstructural changes
during flow are even more complex. At low shear rates, the DEH
theory predicts that the pseudo vector describing the average
alignment of the rods, i.e. the ``director'', undergoes a
continuous ``tumbling'' motion in the plane defined by the
velocity and the velocity gradient vectors. At high shear rates
the director is predicted to align with the
flow\cite{Marrucci89,Larson90}. At intermediate shear rates it is
possible to obtain multiple solutions to the Doi-Edwards-Hess
equation, which are dependent on the initial orientation of the
director~\cite{Faraoni99,Forest03}. For one stable solution called
"wagging" the nematic director oscillates between two angles in
the plane defined by the flow and the gradient of the flow. Other
solutions such as kayaking and log-rolling are also possible, in
which the director oscillates out of the flow-gradient plane at
these intermediate shear rates \cite{Grosso03}.

Experiments on polymeric liquid crystals have confirmed several
predictions of the Doi-Edwards equation. Using a combination of
rheological and rheo-optical measurements it was shown that
nematic solutions of poly(benzyl-glutamate) tumble (PBG) at low
shear rate and become flow aligning at high shear
rates~\cite{Burghardt91}. The existence of a wagging regime and a
potential coexistence of wagging and log-rolling regimes at
intermediate flow rates have also been revealed in
experiments~\cite{Mewis97,Grosso03}. However, there remain
significant difficulties when comparing experiments on polymeric
liquid crystals (PLC) to theoretical predictions. One problem is
that different levels of the microstructure may lead to different
contributions to the stress tensor \cite{Larson96}. In addition to
the molecular contribution to the stress tensor, textural aspects
contribute. The latter include Frank elasticity contributions due
the presence of spatial distortions of nematic director, and
viscous interactions between 'domains'. In addition, there is an
indirect effect to the stress tensor as the defects disturb the
orientation distribution function. These textural contributions to
the total stress dominates the behavior at high concentrations and
low shear rates \cite{Walker96b}, making it difficult to
accurately extract information about the concentration dependence
of different flow transitions. The textural portion of the stress
typically displays scaling of the transient rheological response
with strain rather than with Pe number \cite{Moldenaers86}. The
strain scaling is a typical feature of materials where the time
response is determined by an inherent length scale which in the
case of PLC's is set by the size of the large non-Brownian nematic
domains\cite{Larson91}.

The DEH theory describes the flow behavior of a homogeneous
ensemble of rods but does not consider any polydomain effects.
Therefore an ideal system for testing DEH theory should have small
textural contributions. In this paper we use rod-like \emph{fd}
virus suspensions to access the concentration dependence of the
transition of tumbling to wagging, and wagging to flow aligning.
We show that the contribution of textural stress is very low,
although the spatial distribution of directors still has to be
accounted for. The main motive for using \emph{fd} virus is the
thorough understanding of its equilibrium behavior, which has been
quantitatively described using the Onsager theory extended to take
into account the the semi-flexible nature of \emph{fd} as well as
its surface charge \cite{Purdy03}. Moreover, \emph{fd} has already
successfully been used for (micro-)rheology experiments in the
isotropic phase~\cite{Graf93,Schmidt00}. The aim of the present
paper is to make a comparison between the \emph{dynamic} flow
behavior of \emph{fd} suspensions and the available {\it
microscopic} theoretical predictions of the DEH theory for a
homogeneous system of colloidal rods under shear.

The paper is organized as follows. In section II we discuss the
equation of motion of the orientational distribution function and
the numerical method we use to solve it. The experimental details
about sample preparation and measurements are given in section
III. The results are discussed in five parts: the stationary
viscosity of \emph{fd} suspensions, the concentration and shear
rate dependence of the oscillatory response to a flow reversal,
the relaxation after cessation of flow at high concentration and
\emph{in situ} microscopy under shear. In section IV the textural
contribution to the stress tensor is investigated more detail.
Finally we present a non-equilibrium phase diagram of shear and
concentration dependence of different flow transitions.

\section{Theory}
\label{theory_section}

The distribution of an ensemble of rods can be described by the
probability density function $P(\hat{u}_1,..,\hat{u}_N,\overrightarrow{r}_1,..,\overrightarrow{r}_n)$
of the positions $\{\overrightarrow{r}_i\}$ and orientations $\{\hat{u}_i\}$ of the rods. Ignoring
any spatial correlations, i.e. restricting  to  a monodomain, we have
$P(\hat{u}_1,..,\hat{u}_N,\overrightarrow{r}_1,..,\overrightarrow{r}_n)
=\bar{\rho}P(\hat{u}_1,..,\hat{u}_N)$, where $\bar{\rho}=N/V$ is
the particle density. Therefore, the {\em orientational}
probability density function, or orientational distribution
function (ODF), fully characterizes the system. The time evolution
of the ODF for a suspension of rods during flow is obtained by
solving the equation of motion for the ODF, given by the
$N$-particle Smoluchowski equation.

\begin{eqnarray}
\label{smol} \!\!\!\!\frac{\partial P(\hat{\bf u},t)}{\partial
t}&=&D_r\;\mathcal{\hat{R}}\;\cdot\mbox{\huge{$\{$}}\mathcal{\hat{R}}
P(\hat{\bf u},t)
\nonumber \\
&&\!\!\!\!\!\!\!\!\!\!\!\!\!\!\!\!\!\!\!\!\!\!\!\!\!+
\,DL^2\bar{\rho}P(\hat{\bf u},t)\mathcal{\hat{R}}\;\oint d\hat{\bf
u}'P(\hat{\bf u}',t)|\hat{\bf u}'\times\hat{\bf
u}|\mbox{\huge{$\}$}} \nonumber
\\&& \!\!\!\!\!\!\!\!\! -\mathcal{\hat{R}}\cdot P(\hat{\bf u},t)\hat{\bf u}\times
\left(\hat{\Gamma}\cdot\hat{\bf u}\right),
\end{eqnarray}
where $\mathcal{R}(...)=\hat{u}\times\nabla_{\hat{u}}(...)$ is the
rotation operator with respect to the orientation $\hat{u}$ of a
rod. $D_r$ is the rotational diffusion of a rod at infinite
dilution. Furthermore, $D$ is the thickness of the rods and $L$ is
their length. $\Gamma = \dot{\gamma}\;\hat{\Gamma}$ is the
velocity-gradient tensor with $\dot{\gamma}$ the shear rate. Here
we choose

\begin{eqnarray} \label{gamma}
\hat{\Gamma}\;=\;\left( \begin{array}{ccc} 0&1&0 \\
0&0&0\\
0&0&0
\end{array} \right),
\end{eqnarray}
which corresponds to a flow ${\bf v}$ in the $x$-direction and its
gradient $\nabla{\bf v}$ in the $y$-direction.

The concentration where the isotropic phase becomes unstable
\emph{in the absence of shear flow} can be calculated by solving
the Smoluchowski equation at zero shear rate. This equation agrees
with Onsagers approach to the I-N transition. Often the
Maier-Saupe potential is used instead of the exact potential,
which in fact corresponds to the first term of the Ginzburg-Landau
expansion of the outer product in the exact potential given
between the brackets in Eq. \ref{smol} \cite{Dhont03c}. Under
\emph{flow conditions}, a rich dynamics phase behavior is found as
a function of shear rate and rod concentration. Marrucci and
Maffettone  were the first to solve the equation of motion of the
ODF numerically, restricting themselves to two dimensions in order
to reduce the computational effort\cite{Marrucci89}. They found
that the director undergoes a tumbling motion with respect to the
flow direction, resulting in a negative normal stress $N_1$.
Larson expanded the ODF in three dimensions using spherical
harmonics and truncated the expansion after checking for
convergence \cite{Larson90}. This treatment predicts a transition
from tumbling to "wagging" and finally to flow aligning state with
increasing shear rates. A closure relation is frequently used for
the interaction term on the right side of Eq. \ref{smol}. This can
greatly bias the results, see i.e. Feng et al.\cite{Feng98}. The
location of the flow transitions in the flow-concentration diagram
is very sensitive to the choice of the closure, and  no
satisfactory closure has been found up till now.

In this paper we use a finite element method to numerically solve
the equation of motion for the ODF, thus avoiding the use of any
specific closure relation. As a typical diffusion-convection
equation, this equation describes the diffusive-convective
transport dynamics of an orientation of a homogeneous ensemble of
thin rigid rods. A surface of a sphere is constructed on which a
tip of the rod moves with respect to its center of mass.  The
equation for the probability of finding the tip of a rod in an
area is determined by the transport fluxes on its boundaries due
to (1) the Brownian diffusion (the first term in the brace
brackets of Eq.~\ref{smol}), (2) the convection induced by the
interparticle forces (the second term in the brace brackets of
Eq.~\ref{smol}) and (3) the convection due to the imposed shear
flow (the third term of  Eq.~\ref{smol}).

\begin{figure}
\epsfig{file=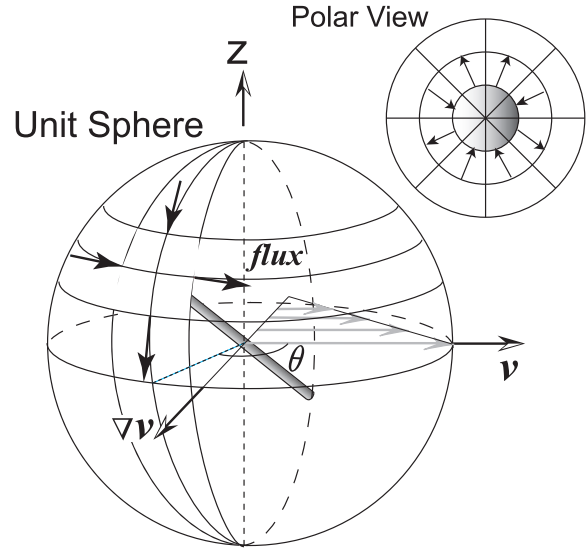,width=3in} \caption{\label{flux} Flux
conservation method used in discritizing Eq.\ref{smol}. The rod
indicates the orientation of the director with respect to the
shear flow. The probability of finding a tip of one rod in the
shaded area of the unit sphere is determined by the flux of the
probabilities through the boundary of that area.}
\end{figure}

To solve Eq.~\ref{smol} numerically, a discretization scheme is
used, and meshes on the surface of a unit sphere are constructed.
For those operators inside the brace brackets which represent the
transport fluxes we apply the central differences approximations.
However, the rotation operator outside of the brace brackets needs
to be discretized using the concept of transport fluxes through
the boundaries of the mesh. In other words, the integral form of
the Eq.(1) is invoked and applied to each of the mesh elements. To
do this the identity, $(\hat{u}\times\nabla_{\hat{u}})
\cdot\mathbf{F}=\nabla_{\hat{u}}\cdot(\mathbf{F}\times\hat{u})$ is
used in order to transform the angular transport flux of a rod to
the translational transport flux of one tip of that rod. It
differs from the conventional method of discretizing a
differential equation where the operators are written explicitly
into the sum of the first- and the second order derivatives and
then the latter are approximated by selected difference schemes.
The advantage of the current method is that, since neighboring
meshes share boundaries, the fluxes leaving one mesh are always
absorbed by the surrounding meshes and vice versa. Therefore,
there is no loss and generation in the total amount of the ODF's
as the computation proceeds (see Fig.\ref{flux}). In practice a
$40 \times 80$ mesh was used on the surface of a unit sphere with
$40$ equi-spaced grids in the polar angle and $80$ equi-spaced
grids in the azimuthal angle in a spherical coordinates. The right
hand side of Eq.\ref{smol} is discritized on the meshes according
to the flux-conservative method mentioned above. A fourth order
Adams' predictor-corrector method~\cite{Korn68} was invoked to
follow the time evolution of the ODF. More details will be
published in a forthcoming paper.

The time-dependent ODF is now used to calculate the
time-dependence of three parameters characterizing the flow
behavior of a nematic phase: (1) $\theta$ describing the angle
between the nematic director and  flow direction, (2) the scalar
magnitude of the director of defined by the order parameter
$P_{2}$ and (3) the total stress of an ensemble of flowing rods.
The angle and magnitude of the order parameter are obtained from
the order parameter tensor
\begin{eqnarray} \label{ordertensor}
{\bf S}=\oint d\hat{\bf u}~\hat{\bf u}\hat{\bf u}P(\hat{\bf u},t).
\end{eqnarray}

The largest eigenvalue of the order parameter tensor $\lambda$,
characterizes the degree of alignment of rods with respect to the
director given by the corresponding eigenvector $\hat{\bf n}$. The
largest eigenvalue of ${\bf S }$ is $1/3$ in the isotropic phase
and to 1 for a perfectly aligned nematic phase. Scalar order
parameter $P_2$ is defined as $P_{2}=(3\lambda -1)/2$.

The stress $\sigma_{12}$ is obtained from the deviatoric part of
the stress tensor derived by Dhont and Briels~\cite{Dhont03c}:
\begin{eqnarray} \label{deviatorica}
{\bf \Sigma}_{D}=\eta_{0}\dot{\gamma}+3\bar{\rho}k_{B}T
\Bigg\{{\bf S}-\frac{1}{3}\hat{\bf I}+\frac{L}{D}\phi {\bf
\Sigma}^{D}_{I}+ \\ \nonumber \frac{1}{6}Pe_{r}[{\bf
S}^{(4)}:\hat{\bf E}-\frac{1}{3}\hat{\bf I}{\bf S}:\hat{\bf E}]
\Bigg\},
\end{eqnarray}
where
\begin{eqnarray} \label{sigmaID}
{\bf \Sigma}^{D}_{I}=\frac{8}{3\pi}\oint d\hat{\bf u}\oint
d\hat{\bf u}' \hat{\bf u}\hat{\bf u}\times\frac{\hat{\bf u}
\times\hat{\bf u}'}{|\hat{\bf u}\times\hat{\bf u}'|}\hat{\bf
u}\cdot\hat{\bf u}' P(\hat{\bf u},t)P(\hat{\bf u}',t)
\end{eqnarray}
and
\begin{eqnarray} \label{S4}
{\bf S}^{(4)}=\oint d\hat{\bf u}~\hat{\bf u}\hat{\bf u}\hat{\bf
u}\hat{\bf u}P(\hat{\bf u},t).
\end{eqnarray}

Here, $\phi=\frac{\pi}{4}D^2L\overline{\rho}$ is the volume
fraction of rods, and $Pe_r=\dot{\gamma}/D_r$ the rotational
P\'{e}clet number which is defined as the shear rate scaled with
the rotational diffusion of a rod at infinite dilution. The first
term between the brackets, ${\bf S}-\frac{1}{3}\hat{\bf I}$, stems
from the Brownian contribution to the stress. The second term
stems from the direct interaction between rods and describes the
elastic contribution to the total stress. The proportionality
constant $\phi \frac{L}{D}$ is the dimensionless rod concentration
and is also called the nematic strength. The terms proportional to
$\sim Pe_r$ stem from the flow of the suspension and described the
viscous contribution to the total stress. This term disappears
instantaneously when the shear is switched off.

\begin{figure}
\epsfig{file=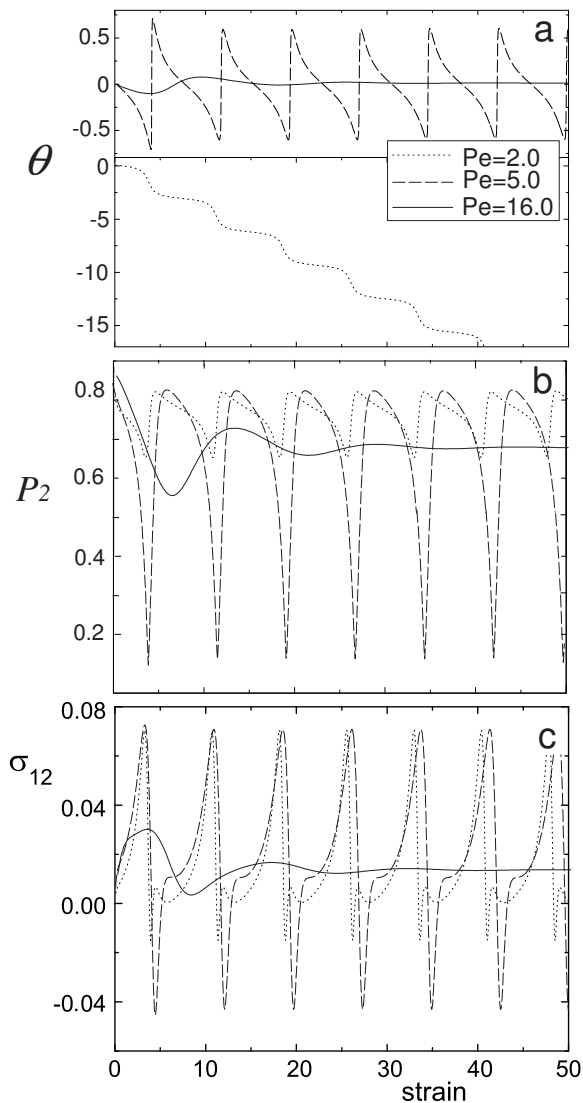,width=3in} \caption{\label{theovector}
Three plots showing the behavior of the angle of the nematic
director $\theta$ (a), the magnitude of the nematic order
parameter(b), and the average stress (c) as a function of strain
after a start up of the flow. The dimensionless rod concentration
is $\phi\frac{L}{D}=4.5$. Data is obtained by numerically solving
Eq.~\ref{smol} using the finite element method. The rods are
initially placed in the flow-gradient plane. For the stress
calculation only the elastic contribution (Eq.~\ref{deviatorica})
was considered.}
\end{figure}

In Fig. \ref{theovector} we plot the evolution of the three
parameters (angle $\theta$, order parameter $P_2$ and stress
$\sigma_{12}$) as a function of strain for different shear rates
at a dimensionless rod concentration of $\phi\frac{L}{D}=4.5$. For
this calculation we used an initial rod orientation in the
flow-gradient plane. The flow behavior between P\'{e}clet numbers
of 4.5 and 5.0 exhibits a sharp transition from tumbling behavior,
where the director continuously rotates, to wagging behavior where
the director hops back and forth between two well defined angles.
At higher shear rates the director is found to be flow aligning.
The order parameter at low shear rates remains unchanged, but is
significantly reduced at the point of the tumbling to wagging flow
transition.

\section{materials and methods}

The viscosity and stress response was measured using an ARES
strain controlled rheometer (TA instruments, Delaware). A double
wall Couette geometry was used because of fairly low viscosity of
the samples. Polarized light microscopy images of {\it fd} under
shear flow were taken using a Linkam CSS450 plate-plate shear
cell.

The physical characteristics of the bacteriophage \emph{fd} are
its length $L= 880\; nm$, diameter $D= 6.6\; nm$, persistence
length of 2200 $nm$ and a charge per unit length of around 10 $e^-
/nm$ at pH 8.2~\cite{Fraden95}. When in solution, \emph{fd}
exhibits isotropic, cholesteric, and smectic phases with
increasing concentration~\cite{Dogic97,Dogic01}. {\it Fd} forms a
cholesteric phase while the DEH theory is valid for nematic
structures. In practice nematic and cholesteric phase are locally
almost identical and the free energy difference between these
phases is very small~\cite{Dogic00}. In this paper we do not
distinguish between these two phases. The {\it fd} virus was
prepared according to standard biological protocols using XL1-Blue
strain of {\it E. coli} as the host bacteria~\cite{Sambrook85}.
The standard yields are approximately $ 50\; mg$ of {\it fd} per
liter of infected bacteria, and virus is typically grown in 6
liter batches. The virus is purified by repetitive centrifugation
(108 000 g for 5 hours) and re-dispersed in a 20 mM Tris-HCl
buffer at pH 8.2.

\subsection{fd as a model hard rod system}\label{fdasmodel}

The Onsager theory for hard rod dispersions predicts a first order
phase transition between a disordered, isotropic phase and an
orientationally ordered, nematic phase. Due to hard core athermal
interactions considered in the Onsager model, the phase diagram is
temperature independent and the rod concentration is the only
parameter that determines the location of the I-N phase
transition. The two points spanning the region of
isotropic-nematic coexistence are called the binodal points. The
spinodal point is located at a rod concentration higher then the
lower binodal point and is determined by the following condition
$\phi \frac{L}{D}=4$. \emph{Fd} viruses are not true hard rods,
due to surface charge and limited flexibility. As a consequence,
their equilibrium phase behavior differs from the ideal hard rod
case described by Onsager based theory, e.g. DEH. The finite
flexibility of \emph{fd} viruses drives the concentration of the
binodal points to a 30\% higher value when compared to equivalent
but perfectly stiff hard rods. In addition, flexibility also
reduces the value of the order parameter of the coexisting nematic
phase. For \emph{fd} the order parameter of the coexisting nematic
is about 0.65 while Onsager theory for hard rods in equilibrium
predicts the order parameter of 0.8~\cite{Purdy03}. The effect of
surface charge is to increase the effective diameter of the rod
$D_{\mbox{\scriptsize eff}}$ and therefore the excluded volume
interaction between charged rods. As a consequence the charge
reduces the real concentration of the phase
transition~\cite{Tang95}.

For the {\it fd} suspension used, the binodal point at high rod
concentration $c_{IN}$ occurs at $11 \;mg/ml$. After taking the
effects of flexibility and charge into account it was shown that
the order parameter of the nematic solution of \emph{fd} is
quantitatively described by the extensions of the Onsager theory
to the semi-flexible case~\cite{Purdy03}. Hence, even though
\emph{fd} is flexible and charged, it can be used to
quantitatively test predictions of the DEH theory. It is, however,
a very difficult and until now unfulfilled task to incorporate
charge and flexibility into a non-equilibrium equation of motion
such as Eq. \ref{smol}. Therefore in this paper we use data from
reference ~\cite{Purdy03} to convert the measured concentration of
\emph{fd} to the nematic order parameter of the sample. After that
we compare experiments and theory at the same values of the order
parameter.

\section{Results}

\begin{figure}
\epsfig{file=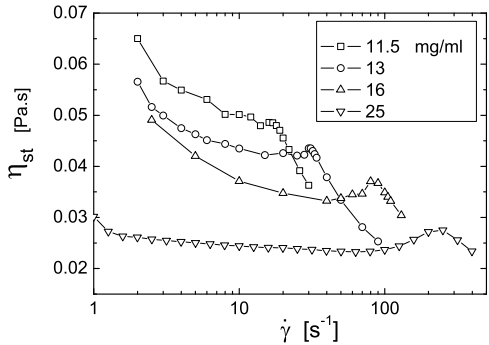,width=3in} \caption{\label{viscosity}
Stationary viscosity as a function of shear rate for four
different concentration of {\it fd} virus at 11.5, 13, 16 and 25
mg/ml.}
\end{figure}

\begin{figure}
\epsfig{file=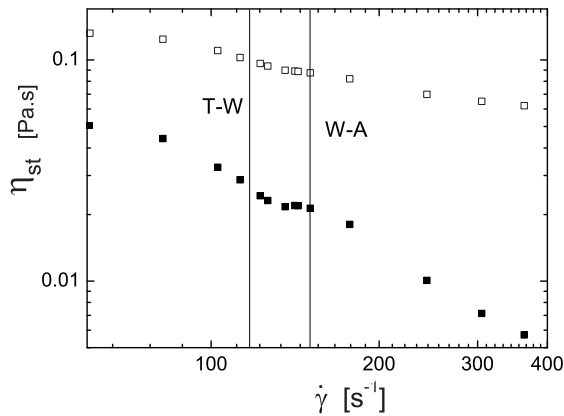,width=3in} \caption{\label{Visccal}
{The theoretical time averaged viscosity at a dimensionless
concentration of $\frac{L}{D}\phi=4.5$ with ($\square$) and
without ($\blacksquare$) the viscous contribution of the rods, as
calculated by solving the equation of motion of the ODF for 20
independent initial orientations of the director. The lines
indicate the transition from tumbling to wagging, and from wagging
to flow aligning as found from Fig. \ref{theovector}. The results
are scaled to the experiment using a typical concentration of
$16;mg/ml$ for $\bar{\rho}$  in Eq. \ref{deviatorica} and
$D^{0}_{\mbox{\scriptsize inf}}=20\; s^{-1}$ for the rotational
diffusion at infinite dilution~\cite{Purdy03,Kramer92}.}}
\end{figure}

\subsection{Stationary viscosity}

The measurements of a stationary viscosity as a function of the
shear rate for different {\it fd} concentrations are shown in Fig.
\ref{viscosity}. For the lowest concentrations of {\it fd} the
viscosity decreases continuously with shear rate except for a
small hesitation at a shear rate of 10 s$^{-1}$. This hesitation
is similar to what is observed for solutions of PBG at low
concentration in solvent {\it m}-cresol
\cite{Kiss78,Moldenaers86}. For {\it fd} at intermediate
concentrations, shear thinning becomes less pronounced, the
hesitation shifts to higher shear rates and turns into a local
maximum. For the highest {\it fd} concentration almost no shear
thinning is observed, only a pronounced peak in the viscosity.
This shear thickening behavior has not been previously reported.

A hesitation in the shear rate dependence of the viscosity was
predicted theoretically by Larson \cite{Larson90}. It was argued
that the transition from the tumbling regime to the wagging regime
implies a broadening of the ODF which leads to higher dissipative
stresses. The broadening of the ODF is illustrated in Fig.
\ref{theovector}b. As can be seen in Fig. \ref{theovector}c it is
not straightforward that ODF broadening really has an effect on
the stress. We calculated the time-dependent viscosity by
numerically solving the equation of motion of the ODF for 20
different initial orientations of the director. From the
time-dependent ODFs we calculated the viscosity using either only
the elastic term or both elastic and viscous terms. The viscosity
is averaged over all 20 traces and a tumbling period after the
transient start up flows have died out. The results are scaled to
the experiment using a typical concentration of $16;mg/ml$ for
$\bar{\rho}$  in eq. \ref{deviatorica} and the value of
$D^{0}_{\mbox{\scriptsize inf}}=20\; s^{-1}$, taken for the
rotational diffusion at infinite dilution~\cite{Purdy03,Kramer92}.
Fig. \ref{Visccal} the stationary viscosity decreases continuously
with increasing shear rate and only shows a hesitation when the
viscous contribution to the stress is not included. The shear rate
where this hesitation occurs corresponds with the shear rate where
the system nematic ordering is significantly reduced and the
transition from tumbling to wagging takes place, as can be
concluded from Fig. \ref{theovector}. Comparing the model
predictions to the experiments it should be noted that the
experimentally observed features are much more pronounced.
Moreover, there is no real reason to leave out the viscous
contribution although it does obscure the behavior we see in the
experiment. Still, the maximum in the viscosity is interpreted as
a signature of the transition from tumbling to a wagging state.

There are three observations to keep in mind when considering
\emph{fd} in the nematic phase under shear flow, which all point
to very low stresses in such systems when compared to polymeric
liquid crystals. First, the viscosity of \emph{fd} in the nematic
phase is two to three orders of magnitude lower than the viscosity
of typical polymeric liquid crystals such as poly(benzyl
glutamate) (PBG) \cite{Vermant94b}, although the difference in
solvent viscosity is only one order of magnitude. Second, the
range over which the viscosity of \emph{fd} suspension varies is
more limited with changing shear rate and rod concentration: the
viscosity lies between 70 times the solvent viscosity for low
shear rate and low rod concentration and 20 times the solvent
viscosity for high shear rate and rod concentration. Moreover, the
viscosity as calculated from the equation of motion of the ODF is
of the same order as the measured viscosity. Third, polymer
nematics exhibit negative first normal stress differences for
certain shear rates as was first observed for PBG
solution~\cite{Kiss78}. This is a direct consequence of the
tumbling of the nematic director. Attempts have been made to
measure the first normal stress difference for nematic {\it fd}
solutions but due to very low force the signals were too small to
be measured.

\subsection{Flow reversal experiments}

\begin{figure}
\epsfig{file=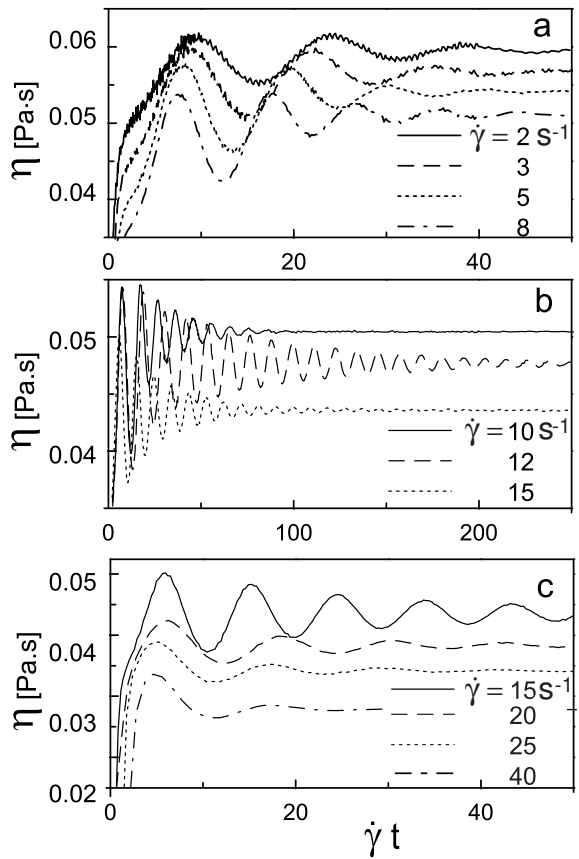,width=3in} \caption{\label{flowrev} The
viscosity of the nematic {\it fd} solution in a response to a flow
reversals. The sample is sheared at shear rate $+\dot{\gamma}$
until the viscosity is equilibrated; at time t=0 the shear rate is
changed to $- \dot{\gamma}$. The concentration of {\it fd} is kept
constant at 11.5 mg/ml. The data can be nicely fitted to Eq.
\ref{fit}. The fits are not shown for clarity. }
\end{figure}

\begin{figure}
\epsfig{file=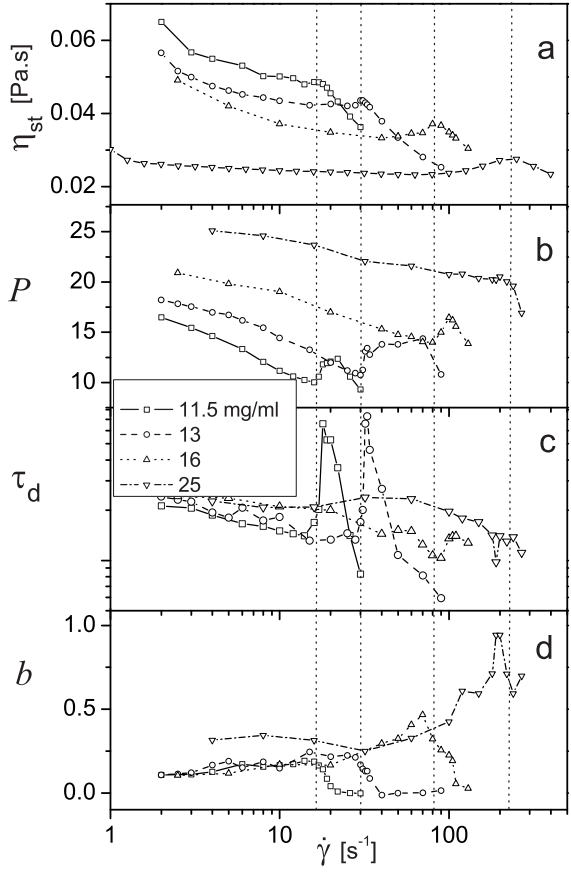,width=3in} \caption{\label{results} a)
Steady state viscosity as function of the shear rate for {\it fd}
virus at four different concentrations. All viscosity curves
exhibit shear thinning at low shear rates followed by a local
maximum in viscosity.  b-d) Behavior of the parameters obtained
from fitting the response of the shear flow reversal experiments
to Eq.~\ref{fit}. The vertical lines indicate the local maximum in
viscosity curves. The local maximum in the steady state viscosity
curve corresponds to maximum of the tumbling period $P$ and
damping constant $\tau_d$ and minimum of asymptotic constat $b$ in
the flow reversal.}
\end{figure}

In flow reversal experiments, the sample is first sheared at a
constant shear rate in one direction until the steady state
condition is reached. Subsequently, the direction of flow is
suddenly reversed while keeping the magnitude of shear rate
constant. Such experiments have been very useful in characterizing
and understanding the dynamics of sheared liquid crystalline
polymers \cite{Moldenaers86}. In the present work, flow reversal
experiments were performed covering a wide range of shear rates
and {\it fd} concentrations. Typical flow reversal experiments are
depicted in Fig. \ref{flowrev} for a {\it fd} concentration of
11.5 mg/ml which corresponds to $c/c_{IN}=1.05$. At the lowest
shear rates a damped oscillatory response is obtained which decays
within few oscillations (Fig. \ref{flowrev}a). Increasing the
shear rate results in a more pronounced oscillatory response,
which damps out relatively slowly. The oscillatory response in
Fig. \ref{flowrev}b is most pronounced at a shear rates of 12
s$^{-1}$. At even higher shear rates, the damping again increases
(Fig. \ref{flowrev}c). In order to quantitatively characterize the
response to a nematic to a flow reversal, the data is fitted to a
damped sinusoidal superimposed onto a asymptotically decaying
function of the following form:
\begin{eqnarray} \label{fit} \eta(t)=\eta_{stat} \cdot\left\{1+A
e^{-\frac{\dot{\gamma}t}{\tau_{d}}}\sin(2\pi\frac{\dot{\gamma}t-\varphi}{P})\right\}(1-b\cdot
g^{\dot{\gamma}t}.)
\end{eqnarray}
This is an empirical choice, but each variable in the fit contains
important information about the behavior of rods in shear flow.
Fig.~\ref{results} shows the behavior of fit parameters as a
function of the shear rate at few selected concentrations of {\it
fd} virus. In this figure we indicate with vertical dashed lines
the shear rates at which the steady state viscosity exhibits a
local maximum for four different concentrations. Interestingly,
these are exactly the same shear rates at which the damping
constant $\tau_d$ as well as the tumbling period $P$ show a sharp
increase. The asymptotic constant $b$, on the contrary, shows a
decrease. These features disappear for the highest {\it fd}
concentration. Presumably the three regions showing different flow
reversal behavior correspond to tumbling, wagging and flow
aligning regime. This will be discussed in more detail in
section~\ref{phase_diagram}. In the next section we first discuss
the concentration and shear rate dependence of the tumbling period
in the regime where rods exhibit tumbling flow behavior.

\begin{figure}
\epsfig{file=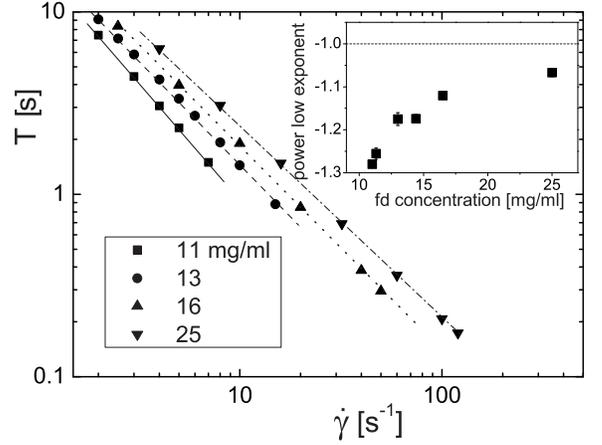,width=3in} \caption{\label{gammapower}
The dependence of the tumbling periods on the shear rate for
different concentrations of the nematic {\it fd}.  The figure
shows that the tumbling period scales with a power low as a
function of the shear rate. The inset shows the power law
dependence of the tumbling period on the shear rate for different
\emph{fd} concentrations.}
\end{figure}

\subsection{Tumbling period as a function of shear rate and rod concentration}

DEH theory predicts that as the De (or Pe) number is increased,
that the 'molecular' period of oscillation decreases with
increasing shear rate in the tumbling regime~\cite{Larson90}. This
feature was never fully explored, since in most polymeric liquid
crystals it was found that the tumbling period was strain scaling,
implying that the response overlaps when the period is scaled with
the applied shear rate and the stress is normalized by its steady
state value. The strain scaling arises as a consequence of the
presence of a large, non-Brownian, length scale in the sample that
determines the time response, even at relatively high De (or Pe)
numbers. This most probably is the domain size characterizing the
nematic texture. The log-log plot of the tumbling period
(T=P/$\dot{\gamma}$) as a function of the shear rate is shown in
Fig.~\ref{gammapower}. Here the data are only shown for a low
shear rate region which is associated with the tumbling region.
Strain scaling, if present, would give a slope of -1. However, as
can be seen in the inset of Fig.\ref{gammapower}, the reciprocal
indicating strain scaling is only approached and not reached at
the highest rod concentration studied here.

\begin{figure}
\epsfig{file=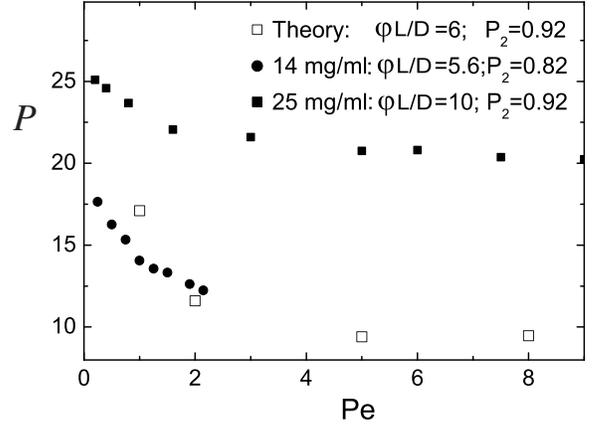,width=3in} \caption{\label{Pvsrate} {The
period of the oscillations (in units of strain) as a function of
the P\'{e}clet number, where the shear rate is scaled with the
rotational diffusion of \emph{fd} at infinite dilution. }}
\end{figure}

The shear rate dependence of the tumbling period is compared to
the theoretical prediction for the same rod concentration as well
as the same order parameter, see Fig.~\ref{Pvsrate}. The reason
for using the order parameter to assess the theoretical
predictions was discussed at length in section~\ref{fdasmodel}.
For purposes of comparison the order parameter was obtained form
x-ray experiments and the value of $D^{0}_{\mbox{\scriptsize
inf}}=20\; s^{-1}$. We emphasize that DEH theory is microscopic
and that there are no adjustable parameters in the comparison
between theory and experiments. Clearly there is a qualitative
correspondence between theory and experiment, both showing a
continuous decrease of the period. The quantitative
correspondence, on the other hand, is limited. This is probably
due to fact that texture, although not dominating the response, is
still present. It will be shown later in
section~\ref{phase_diagram} that the shear rate and rod
concentration dependence of a tumbling to wagging and wagging to
flow-aligning transition agree much better with DEH theory.

\begin{figure}
\epsfig{file=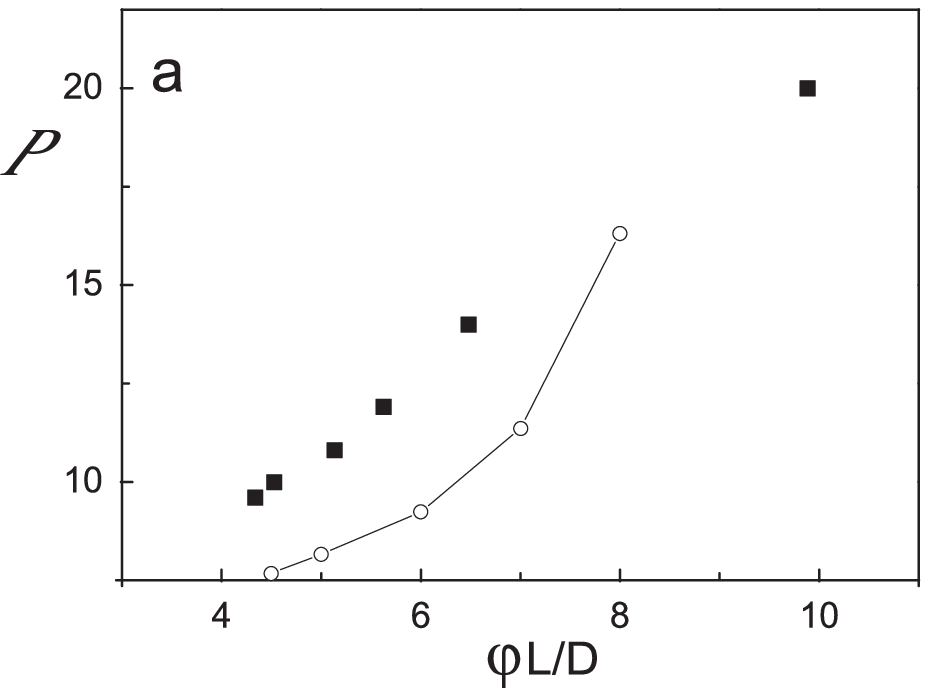,width=2.9in}
\epsfig{file=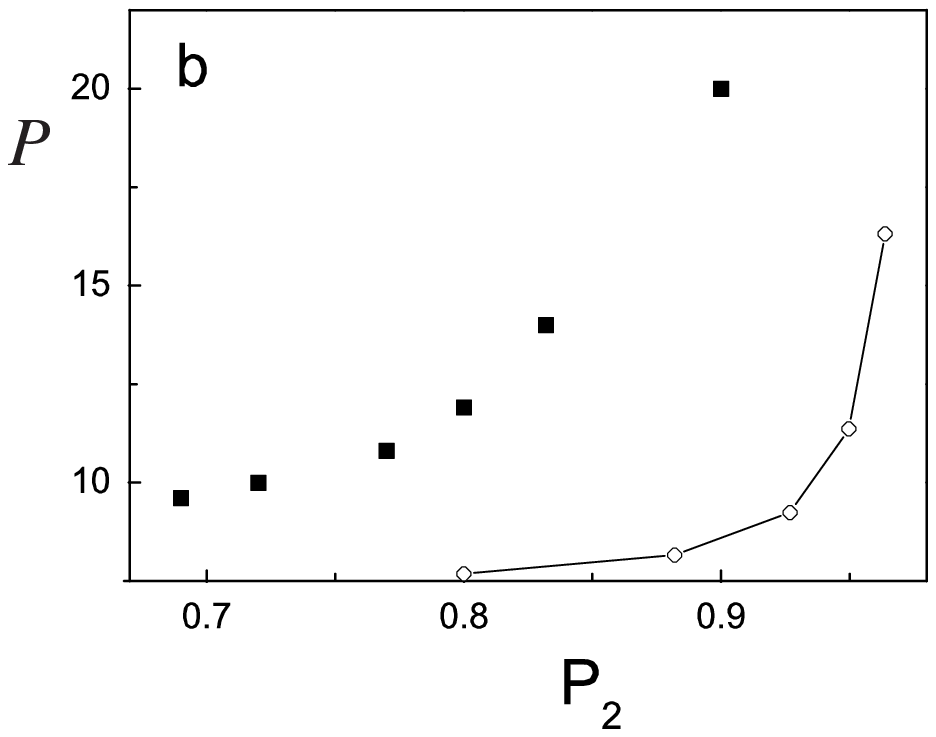,width=2.9in} \caption{\label{PvsC} {The
period of the oscillations (in units of strain) as a function of
the dimensionless concentration (a) and order parameter (b). The
shear rate was chosen at the point we identify with the tumbling
to wagging transition for experiment ($\blacksquare$) and exact
theory ($\circ$).}}
\end{figure}

The concentration dependence of the tumbling period is shown in
Fig.~\ref{PvsC}. Here, theory and experiments are compared at a
fixed shear rate at which the tumbling to wagging flow transition
occurs. The tumbling period increases with increasing rod
concentration (Fig.~\ref{PvsC}a) or, equivalently, increasing
order parameter of the nematic phase (Fig.~\ref{PvsC}b). The
increase of the tumbling period with increasing order parameter
was already predicted using a linearized version of the DEH
theory\cite{Kuzuu84} .

In conclusion, the absence of strain scaling of the tumbling
period and the qualitative agreement between theory and experiment
the tumbling period indicates that the response of the suspension
of \emph{fd}-virus is dominated by the molecular elasticity
arising from the distortion of the ODF of particles.

\subsection{Relaxation at high concentration}

In order to measure the relative magnitude of the elastic texture
contribution to the overall stress, relaxation experiments were
performed. For polymeric liquid crystals like
poly\-benzyl\-glutamate (PBG) solutions in m-cresol, Walker et al.
\cite{Walker96b} showed that there are three different regimes of
relaxation behavior, each of which is related to a distinct
structural relaxation. There is a "fast" relaxation of the nematic
fluid; a "slower" relaxation that exhibits scaling with the shear
rate before the cessation of flow, which is due to the indirect
contribution of the texture to the overall stress; and a
"long-time" relaxation due to the reorganization of the texture on
a supra-molecular level which will not be addressed here.

\begin{figure}
\epsfig{file=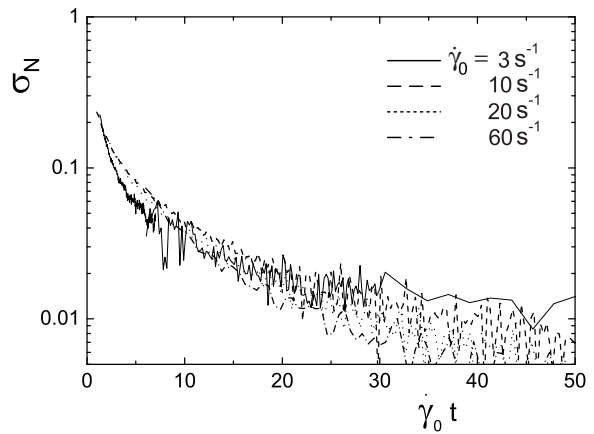,width=3in}
\caption{\label{relaxation} {The stress relaxation after cessation
of flow for \emph{fd} at $25\; mg/ml$ ($c/c*=2.3$), varying the
initial shear rate. The time is scaled by the initial shear rate.
The stress is normalized by the stress before the cessation of
flow.}}
\end{figure}

Stress relaxation experiments were performed in the low  shear
rate ``tumbling'' region, at shear rates smaller than those
corresponding to the maximum in the viscosity. The sample used had
a relatively high \emph{fd} concentration of $25\; mg/ml$,
corresponding with $c/c_{IN}=2.3$. Some typical responses to the
cessation of flow are depicted in Fig. \ref{relaxation}. The
stress is normalized to its value before the cessation of flow,
and the time axis is scaled by the shear rate. The fast component
of the decay takes place at less than a tenth of a second, which
is comparable to the response of the force re-balanced transducer
and therefore not shown. The slow component of the stress
relaxation scales when time is multiplied with the previous shear
rate, but only from the point that the stress has decayed to less
than 30~\% of its original value, or less for higher initial shear
rates. From Fig. \ref{relaxation} it can be concluded that the
contribution to the stress for the highest concentration used and
for low shear rates is 30~\%. This is the absolute upper limit for
the samples used in this paper. It should be noted that for PBG
solutions 30 \% it was found to be the lower limit
\cite{Walker96b}.

\subsection{\emph{In situ} microscopy}

\begin{figure}
\epsfig{file=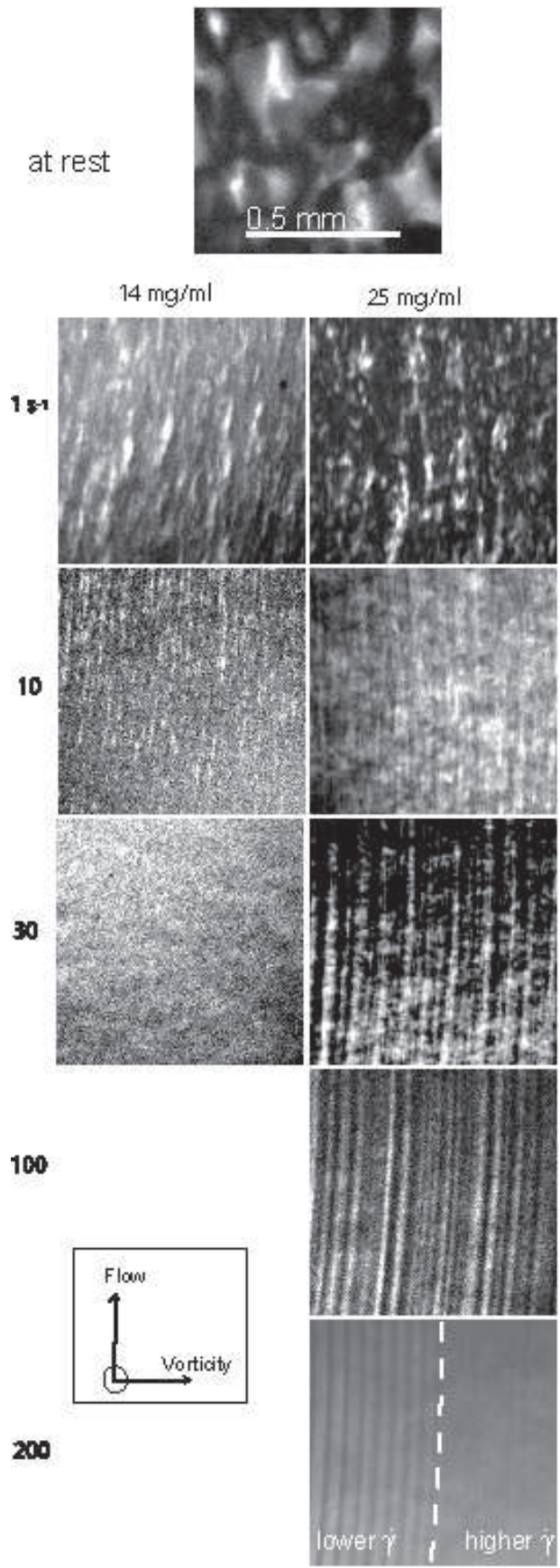,width=0.45\textwidth}
\caption{\label{shearstruc} The polarization images of the nematic
{\it fd} at 14 and 25 mg/ml for a range of different steady state
shear rates. The dashed line in the bottom right image indicates
the border between the structured and unstructured regions.}
\end{figure}

The flow-induced changes of the liquid crystalline texture during
steady state shear flow were studied using a plate-plate geometry
in combination with a polarization microscope. Measurements were
performed for \emph{fd} concentrations of 14 mg/ml and 25 mg/ml.
Typical images are shown in Fig. \ref{shearstruc} for different
shear rates. Interestingly the characteristic size of the
"domains" was very large. Birefringent regions of up to half a
millimeter were observed under static conditions. When the sample
is subjected to shear flow, these domains will elongate and
eventually disappear, at values of the shear rate which correspond
to the maximum in the viscosity (see Fig.\ref{results}a). An
important difference between the two concentrations is that the
elongated domains merge into bands for high rod concentration,
whereas for the low concentration the structure disappears before
such bands are formed. Interestingly, this transition to a banded
structure in the high concentration fluid takes place at a shear
rate which is higher than the shear rate where the low
concentration fluid loses its features.

\section{Discussion}

When comparing the flow behavior of the polymeric nematic phase
and the colloidal nematic phase of the dispersed \emph{fd}
viruses, the most striking observation is the qualitative
agreement between the two systems, despite the fact that \emph{fd}
is an order of magnitude larger. The viscosity of the {\it fd}
nematic is much smaller and the rotational diffusion of \emph{fd}
is much slower when compared to polymeric liquid crystals. Flow
reversal experiments reveal typical transitions in the transient
rheological behavior: damped oscillations occur at low shear rates
changing to undamped oscillations at intermediate shear rate,
which disappear if the shear rate is increased even further; the
time scale of the oscillations of the stress transients is
comparable. Also other well known phenomena like the formation of
very large bands upon cessation of flow along the vorticity
direction which have been studied in detail in polymeric systems
\cite{Vermant94}, can also be observed here (data not shown).

Having established that \emph{fd} virus dispersions indeed undergo
a tumbling motion under flow, the \emph{dynamic} behavior of
\emph{fd} suspensions can be rationalized on the basis of the
microscopic theoretical predictions for a homogeneous system of
rods under shear. Doing so, one important prerequisite needs to be
fulfilled, namely that the dominating contribution to the stress
is coming from the nematic fluid and not from the texture. It will
be argued here that this indeed is the case. Having done so, we
will be able to map out a phase diagram of the dynamic transitions
from tumbling to wagging to flow aligning.

\subsection{Textural evolution during flow}

The word ``texture'' refers to disclination points and lines where
the director of the nematic phase changes discontinuously, marking
domains in the sample. When a system containing these domains and
disclinations is subjected to shear flow, part of the dissipated
energy is used to destroy these structures. Fig.~\ref{shearstruc}
shows that the domains tend to elongate and align with the flow.
Disclinations can also cause a direct contribution to the total
stress resulting in a high viscosity and a very pronounced shear
thinning behavior, typically referred to as region I
~\cite{Walker94}. Experiments on polymeric liquid crystals have
revealed several features of the flow behavior of nematic liquid
crystals which are attributed to the presence of texture in the
nematic phase. Tumbling induces distortions in the director field
and the defects arrest the tumbling, thereby inducing an elastic
stress. The length scale over which this distortion occurs, i.e.
the 'domain' length scale,  is an inherent non-Brownian length
scale, see ref. \cite{Burghardt90}. As a consequence, stress
patterns during flow reversal will display strain scaling. Also
the damping of the oscillations is explained on the base of the
presence of the polydomain structure, where e.g. the "friction"
between the domains would lead to a damping of the
oscillations~\cite{Larson91,Kawaguchi99}. The scaling of the
stress relaxation process after the flow is stopped with shear
rate has been explained using the same arguments. From such an
experiment the relative contribution to the total stress of a
homogeneous nematic phase and the polydomain texture can be
estimated since the relaxation dynamics of the nematic phase is
much faster than that of polydomain structure~\cite{Walker96b}.

The micrographs in Fig.~\ref{shearstruc} clearly reveal that
texture under flow exists in nematic \emph{fd} dispersions. Their
contribution to the rheology is, however, far less prominent when
compared to polymeric liquid crystals such as PBG. This we can
infer from several observations. First, very moderate shear
thinning is observed in the low shear rate regime for the
\emph{low} concentrations, which gradually disappears with
increasing concentration (Fig. \ref{results}b). This is very
similar to theoretical predictions for a homogeneous nematic phase
(Fig. 2b in Ref.~\cite{Dhont03c}). Also, the calculated and
measured viscosity are of the same order of magnitude. In
contrast, shear thinning can be fairly strong in the low shear
rate region (Region I) where texture dominates the response and it
will increase with increasing concentration~\cite{Marrucci93},
although also other microstructural features can contribute here
\cite{BurghardtRI}. Second, the tumbling period is not strain
scaling (Fig. \ref{gammapower}), which could be due to either a
smaller relative magnitude of the textural stress or due to the
fact that we are not in a low enough Pe regime. Third, 'strain'
scaling is recovered for the slow 'textural' relaxation process
after the flow has been stopped. This experiment shows that at the
highest rod concentrations used and at low shear rates the
distortional textural contribution is about 30~\%. For most
experiments done this value is probably significantly lower. So,
where texture is important, even dominating the  stress response
for molecular LCPs, molecular elasticity is far more dominating
the \emph{fd}-virus. Though we just argued that the texture does
not dominate the shear response of the system, this does not mean
that the shear response is not influenced by texture. For one the
oscillations we observe are still strongly damped, and the damping
only decreases when the transition to the flow aligning state is
reached (see the behavior of $\tau_d$ in Fig. \ref{results}).
Moreover, the presence of texture might explain the discrepancy in
the behavior of the period of the oscillations between experiment
and theory (Fig. \ref{PvsC}). Most importantly, we know from
microscopy that texture is present under shear (see Fig.
\ref{shearstruc}). It should be noted, however, that the size of
the polydomain structure of the \emph{fd} dispersions is one order
of magnitude bigger as compared to PBG~\cite{Vermant94b}, so that
the density of disclination lines and points is about three orders
of magnitude lower for \emph{fd}. Note that the length scale of
the texture during flow is still small compared to the dimension
of the flow cell. Since the contribution of texture scales with
the density of the disclinations~\cite{Marrucci93}, texture will
be far more dominating for e.g. PBG than for \emph{fd}, even when
elastic constants are almost the same for the two
systems~\cite{Dogic00},\cite{taratuta85}).

\subsection{Phase diagram of dynamical flow transitions}
\label{phase_diagram}

In this section the experimental results or combined and a
non-equilibrium phase diagram of {\it fd} rods under shear flow is
presented. The results for the four fit parameters plotted in
Fig.~\ref{results} show clear transitions at well defined shear
rates for all {\it fd} concentration. Although they only give an
indirect proof of the transitions, they can be used to infer
information about the flow transitions. For all {\it fd}
concentrations (except for the highest one) the shear rate where
the maximum viscosity is reached is identical with the shear rate
where the period as well as the damping constant start to increase
(indicated by the vertical dashed lines in Fig. \ref{results}).
The microscopic observations are in fairly good agreement with the
transitions inferred from the rheology. Upon approaching the
tumbling to wagging transition from tumbling to flow aligning, the
texture becomes to faint to resolve in the microscope and texture
subsequently disappears upon reaching the FA region. For the high
{\it fd} concentration, i.e. the sample showing shear banding
(Fig.~\ref{shearstruc} last), one can identify a sharp transition
from a structured to an unstructured region in the same
micrograph. Since this picture was taken in the plate-plate
geometry, there is a shear rate distribution across the image: the
shear rate is increasing going from the left side to the right. A
sharp spatial transition therefore also represents a sharp
transition at a given shear rate. Although, due to the method of
zero gap-setting, the value of the shear rate is not exactly known
($\pm 20 \%$), one can still identify the shear rate where
structure disappears as the shear rate where the viscosity reaches
its local maximum (the down pointing triangles in
Fig.~\ref{results}a). For low {\it fd} concentration of ($14\;
mg/ml$) the structure disappears around the point where the
viscosity reaches its local maximum, although the morphological
transition for the lower concentration is less abrupt.

\begin{figure}
\epsfig{file=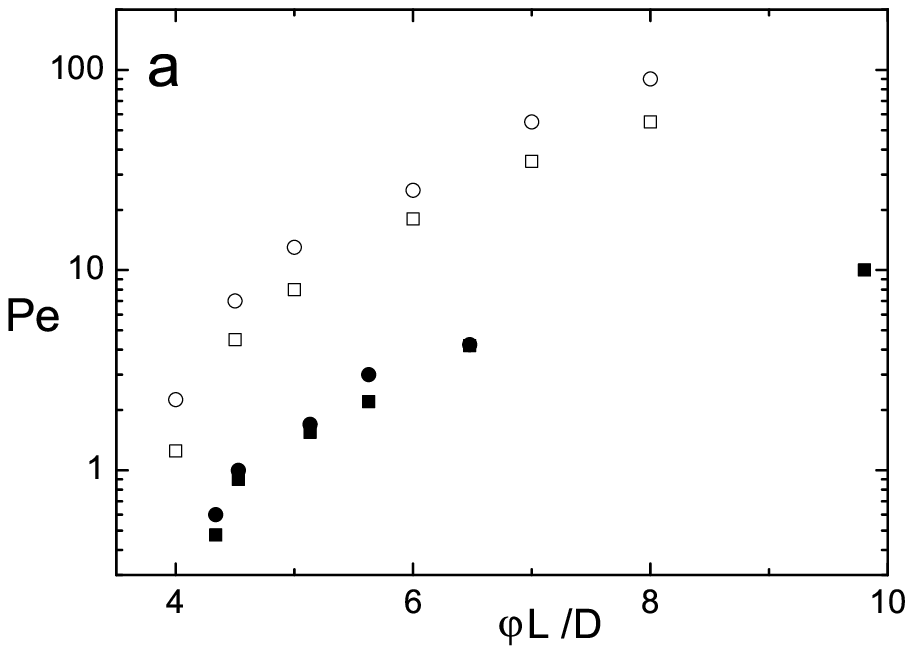,width=3.5in}
\epsfig{file=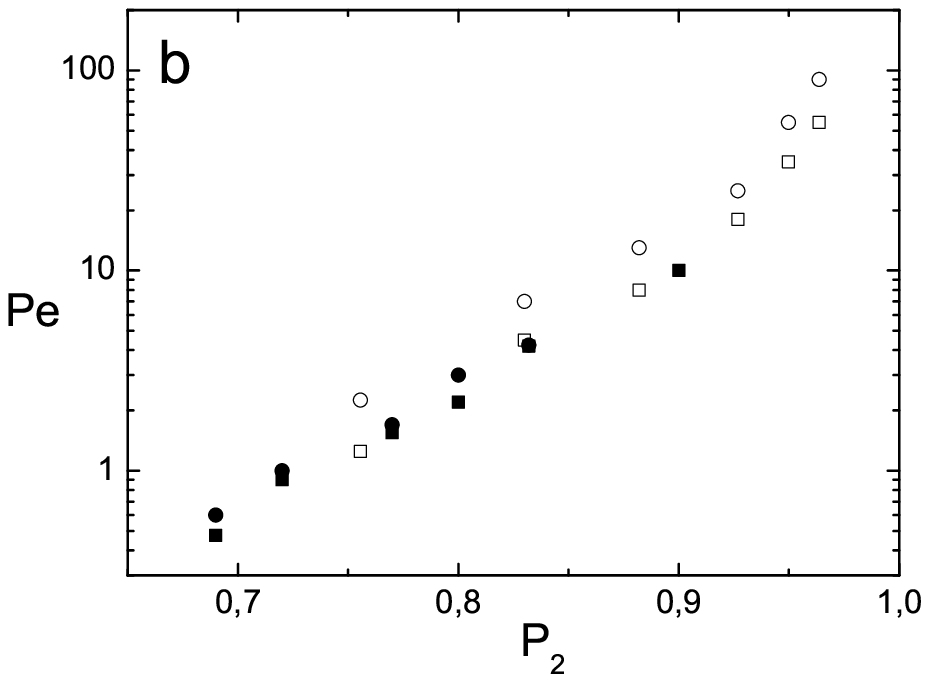,width=3.5in}
\caption{\label{cvsg}  {Phase diagram of flow transitions for the
nematic {\it fd} phase as a function of dimensionless
concentration (a) and order parameter (b). The experimental points
indicate the P\'{e}clet numbers where the viscosity shows a local
maximum ($\blacksquare$) and where the damping constant $\tau_d$
($\bullet$) reaches a maximum. The theoretical points indicate
tumbling to wagging ($\square$) and wagging to flow aligning
($\circ$) transitions.}}
\end{figure}

Fig. \ref{cvsg} shows the behavior of flow transitions as a
function of shear rate for various {\it fd} concentrations. For
the experiment we plotted the P\'{e}clet numbers where the
viscosity shows a local maximum and where the damping constant
reaches a maximum. The theoretical predictions for the tumbling to
wagging and wagging to flow aligning transitions are obtained from
the plots the angle of the nematic director $\theta$ under flow,
see Fig. \ref{theovector}. Similar to the method used in
Fig.~\ref{PvsC}, the experimental concentration is scaled to the
theoretical concentration in two different ways: effective
concentration (Fig. \ref{cvsg} a) and the order parameter $P_2$
(Fig. \ref{cvsg} b).  This figure was shown in a preliminary paper
without a detailed explanation \cite{Lettinga04a}. The shear rate
is rescaled to the P\'{e}clet number by using the rotational
diffusion coefficient at infinite dilution. Fig. \ref{cvsg} allows
us to draw some important conclusions. First, it is clear that
scaling the concentration with the equilibrium order parameter
gives better agreement when compared to the scaling by the
dimensionless concentration. The fact that theory and experiment
agree without using any fitting parameters ($P_2$ was obtained in
a separate experiments~\cite{Purdy03}) leads to the conclusion
that the DEH theory describes the flow behavior of the {\it fd}
nematics quite well, as long as the effects of flexibility and
charge of the experimental rods are included in the calculation of
the order parameter. Less convincing agreement is obtained when
comparing the experimental and theoretically calculated periods
(Fig. \ref{PvsC}). The reason for this could be the remaining
textural contribution to the overall stress which, although small,
cannot be neglected. Since we deduce from Fig. \ref{cvsg}b, that a
dimensionless concentration of $\phi \frac{L}{D}=4$ corresponds
with a \emph{fd} concentration of $16\; mg/ml $, we used this
number \emph{a posteriori} to scale the calculated molecular
viscosity in Fig. \ref{Visccal}. The pure elastic contribution
shows a very nice quantitative correspondence with the
experimental data. Interestingly, when the viscous term is added,
the theoretical viscosity is higher than the experimental
viscosity, despite of the fact that no hydrodynamics is
incorporated.

In the previous subsection it was argued that the influence of
textural contribution to the stress tensor of {\it fd} are
relatively small, as compared to PLC's. There are however strong
indications that the dynamic behavior is influenced by the
macroscopic bands which are formed for the samples at the highest
concentrations used (see Fig. \ref{shearstruc} end). As can be
seen in Fig. \ref{results}, the typical features for the
transition to wagging disappear: there is no increase in the
damping constant, nor in the period of the oscillations. Moreover,
the theory shows only a moderate hesitation of the stationary
viscosity (which even disappears then the viscous term is added,
Fig. \ref{Visccal}), whereas in experiments a local peak is
observed which is more pronounced with increasing concentration.
The microscopy pictures results that at high concentrations the
systems finds another way to handle the distortion of the particle
distribution at high shear rates by forming shear bands where the
overall orientational distribution is alternating, as was already
observed and partially explained for the polymeric systems
\cite{Larson93a,Larson93b,Vermant94b}. In the present work, the
concentration dependence of the phenomenon at hand suggests that
this merits further experimental as well as theoretical work. In
this context one should not forget that we compar experiments on
charged and semi-flexible \textit{fd} with theory for hard and
stiff rods. It could well be that these factors also play an
important role. It will be a major challenge especially to take
the semi-flexibility into account in the equation of motion.

\section{Conclusions}

Colloidal suspensions of rod-like \emph{fd} viruses are an ideal
model system to study the behavior of the nematic liquid
crystalline phase under shear flow. Flow reversal experiments show
signatures for tumbling, wagging, and flow aligning behavior, very
similar to the behavior found in polymeric liquid crystals.  The
rigid rod nature of the \emph{fd} suspension, possibly combined
with a smaller relative textural contributions to the overall
stress tensor make \emph{fd}-virus a suitable model system for the
DEH theory. Important in this respect is that the overall
viscosity is only one to two orders of magnitude higher than the
solvent viscosity. Also it is important to note that stress
relaxation experiments combined with the absence of strain scaling
in flow reversal experiments suggest that there is only a limited
contribution of textural aspects to the overall stress, even for
the highest {\it fd} concentration used in this work. The shear
thickening of the viscosity observed for a range of {\it fd}
concentrations is as yet, unexplained. The maximum in the
viscosity occurs at the critical shear rate where the tumbling to
wagging transition takes place. Microscopic observations show that
at this shear rate the morphological features disappear,
suggesting a strong connection between the dynamic transitions and
structure formation.

The experimental results have been compared to a microscopic
theory for rod like molecules subjected to shear flow. A
non-equilibrium phase diagram is constructed, describing the
transitions from tumbling to wagging and from wagging to
flow-aligning as a function of rod concentration and applied shear
stress. When scaling the results to the concentration where the
isotropic-nematic transition takes place, the experiment and
theory show only a qualitative agreement, possibly due to the fact
that the real rods are are both semi-flexible and charged.
However, when scaling the results using the order parameter, which
is determined by the interactions between the rods, theory and
experiment show an excellent agreement without using any fit
parameters. Thus, it can be concluded that the Doi-Edward-Hess
theory accurately captures the dynamic features of a hard rod
system. \emph{Fd} dispersions constitute such a hard rod system as
long as flexibility and charge are properly taken into account,
which can be simply achieved by using the order parameter to scale
the data. More theoretical work is needed, however, to explain the
clear connection between the observed band formation at high
concentrations and the dynamic transitions, and to incorporate the
effect of flexibility of the rods.

\section*{Acknowledgement}
We thank Jan Dhont for many discussions and critical reading of
the manuscript. Pier-Luca Maffetonne is acknowledged for
stimulating remarks. MPL is supported by the Transregio SFB TR6,
"Physics of colloidal dispersions in external fields". ZD is
supported by Junior Fellowship at Rowland Institute at Harvard.
The authors acknowledge support of the EU (6th FP) in the
framework of the Network of Excelence  "SOFTCOMP".


\begin{thebibliography}{10}

\bibitem{Larson96} R. G. Larson.
\newblock On the relative magnitudes of viscous, elastic and texture
stresses in liquid crystalline PBG solutions
\newblock {\em  Rheol. Acta}, 35 (2): 150-159, 1996.

\bibitem{BurghardtRI} Ugaz V. M. and Cinader D. K. and W. R. Burghardt
\newblock Origins of region I shear thinning in model lyotropic liquid crystalline polymers
\newblock {\em  Macromolecules }, 30 (5): 1527-1530, 1997.


\bibitem{Burghardt90}
W.~R. Burghardt and G.~G. Fuller.
\newblock Transient shear flow of nematic liquid crystals: Manifestations of
  director tumbling.
\newblock {\em J. Rheol.}, 34(6):959--992, 1990.

\bibitem{Burghardt91}
W.~R. Burghardt and G.~G. Fuller.
\newblock Role of director tumbling in the rheology of polymer liquid crystal
  solutions.
\newblock {\em Macromolecules}, 24:2546, 1991.

\bibitem{Dhont03c}
J.~K.~G. Dhont and W.~J. Briels.
\newblock Viscoelasticity of suspensions of long, rigid rods.
\newblock {\em Colloid Surface A}, 213(2-3):131--156, 2003.

\bibitem{Dogic97}
Z.~Dogic and S.~Fraden.
\newblock Smectic phase in a colloidal suspension of semiflexible virus
  particles.
\newblock {\em Phys. Rev. Lett.}, 78:2417, 1997.

\bibitem{Dogic00}
Z.~Dogic and S.~Fraden.
\newblock Cholesteric phase in virus suspensions.
\newblock {\em Langmuir}, 16:7820--7824, 2000.

\bibitem{Dogic01}
Z.~Dogic and S.~Fraden.
\newblock Development of model colloidal liquid crystals and the kinetics of
  the isotropic-smectic transition.
\newblock {\em Phil. Trans. R. Soc. Lond. A.}, 359:997, 2001.

\bibitem{Doi86}
M.~Doi and S.~F. Edwards.
\newblock {\em The Theory of Polymer Dynamics}.
\newblock Oxford, 1986.

\bibitem{Faraoni99}
V.~Faraoni, M.~Grosso, S.~Crescitelli, and P.~L. Maffettone.
\newblock The rigid-rod model for nematic polymers: An anaysis of the shear
  flow problem.
\newblock {\em J. Rheol.}, 43:829, 1999.

\bibitem{Feng98}
J.~Feng, C.~V. Chaubal, and L.~G. Leal.
\newblock Closure approximations for the doi theory: Which to use in simulating
  complex flows of liquid-crystalline polymers?
\newblock {\em J. Rheol.}, 42(5):1095--1119, 1998.

\bibitem{Forest03}
M.~G. Forest and Q.~Wang.
\newblock Monodomain response of finité-aspect-ratio macromolecules in shear
  and realted linear flows.
\newblock {\em Rheol. Acta}, 42:20--46, 2003.

\bibitem{Fraden95}
S.~Fraden.
\newblock {\em Observation, Prediction, and Simulation of Phase Transitions in
  Complex Fluids}, volume 460 of {\em NATO-ASI - Series C}, pages 113--164.
\newblock Kluwer Academic Publishers, Dordrecht, m. baus and l. f. rull and j.
  p. ryckaert, edition, 1995,.

\bibitem{Graf93}
C.~Graf, H.~Kramer, M.~Deggelmann, M.~Hagenb\"{u}chle, Ch. Johner,
Ch. Martin,
  and R.~Weber.
\newblock Rheological properties of suspensions of interacting rodlike fd-virus
  particles.
\newblock {\em J. Chem. Phys.}, 98(6):4921--4928, 1993.

\bibitem{Grosso03}
M.~Grosso, S.~Crescitelli, E.~Somma, J.~Vermant, P.~Moldeaers, and
P.~L.
  Maffettone.
\newblock Prediction and observation of sustained in a shear liquid crystalline
  polymer.
\newblock {\em Phys. Rev. Lett.}, 90:098304, 2003.

\bibitem{Hess76}
S.~Hess.
\newblock Fokker-planck-equation approach to flow alignment in liquid crystals.
\newblock {\em Z. Naturforsch.}, 31(a):1034--1037, 1976.

\bibitem{Kawaguchi99}
M.~N. Kawaguchi and M.~M. Denn.
\newblock A mesoscopic theory of liquid crystalline polymers.
\newblock {\em J. Rheol.}, 43(1):111--124, 1999.

\bibitem{Kiss78}
G.~Kiss and R.~S. Porter.
\newblock Rheolog of concentrated solutions of poly($\gamma$-benzyl-glutamate).
\newblock {\em J. of Polym. Sci.}, 65:193--211, 1978.

\bibitem{Korn68}
Korn and Korn.
\newblock {\em Mathematical Handbook for Scientists and Engineers}.
\newblock Mc-Graw and Hill, 1968.

\bibitem{Kramer92}
H.~Kramer, M.~Deggelmann, C.~Graf, M.~Hagenbtichle, C.~Johner, and
R.~Weber.
\newblock Electric birefringence measurements in aqueous fd virus solutions.
\newblock {\em Macromolecules}, 25:4325--4328, 1992.

\bibitem{Kuzuu84}
N.~Kuzuu and M.~Doi.
\newblock {\em J. Phys. SOC. Jpn.}, 53:1031, 1984.

\bibitem{Larson90}
R.G. Larson.
\newblock Arrested tumbling in shearing flows of liquid crystal polymers.
\newblock {\em Macromolecules}, 23:3983--3992, 1990.

\bibitem{Larson93b}
R.G. Larson.
\newblock Roll-cell instabilities in shearing flows of nematic polymers.
\newblock {\em Liquid Crystals}, 37(2):175--197, J. Rheol.

\bibitem{Larson91}
R.G. Larson and M.~Doi.
\newblock Mesoscopic domain theory for textured liquid crystalline polymers.
\newblock {\em J. Rheol.}, 35(4):539--563, 1991.

\bibitem{Larson93a}
R.G. Larson and D.~W. Mead.
\newblock The ericksen number and deborah number cascades in sheared polymeric
  nematics.
\newblock {\em Liquid Crystals}, 5(2):151--169, 1993.

\bibitem{Lettinga04a}
M.P. Lettinga and J.K.G. Dhont.
\newblock Non-equilibrium phase behavior of rod-like viruses under shear flow.
\newblock {\em J. Phys.: Condens. Matter 16 (2004)}, 16:S3929–S3939, 2004.

\bibitem{Marrucci93}
G.~Marrucci and F.~Greco.
\newblock Flow behavior of liquid crystaline polymers.
\newblock {\em Adv. Chem. Phys.}, 86:331, 1993.

\bibitem{Marrucci89}
G.~Marrucci and P.~L. Maffettone.
\newblock Description of the liquid-crystallin phase at high shear rates.
\newblock {\em Macromolecules}, 22:4076, 1989.

\bibitem{Mewis97}
J.~Mewis, M.~Mortier, J.~Vermant, and P.~Moldenaers.
\newblock Experimental evidence for the existence of a wagging regime in
  polymeric liquid crystals.
\newblock {\em Macromolecules}, 30(5):1323--1328, 1997.

\bibitem{Moldenaers86}
P.~Moldenaers and J.~Mewis.
\newblock Transient behavior of liquid crystalline solutions of
  poly(benzylglutamate).
\newblock {\em J. Rheol.}, 30(1):567--584, 1986.

\bibitem{Onsager49}
L.~Onsager.
\newblock The effect of shape on the interaction of colloidal particles.
\newblock {\em Annals of the New York academy of science}, 51:62--659, 1949.

\bibitem{Purdy03}
K.~R. Purdy, Z.~Dogic, S.~Fraden, A.~R\"{u}hm, L.~Lurio, and
S.~G.~J. Mochrie.
\newblock Measuring the nematic order of suspensions of colloidal fd virus by
  x-ray diffraction and optical birefringence.
\newblock {\em Phys. Rev. E.}, 67:031708, 2003.

\bibitem{Sambrook85}
J.~Sambrook, E.~F. Fritsch, and T.~Maniatis.
\newblock {\em Molecular Cloning: A Laboratory Manual}, chapter~4.
\newblock Cold Spring Harbor Laboratory Press, 2 edition, 1989.

\bibitem{Schmidt00}
F.~G. Schmidt, B.~Hinner, E.~Sackmann, and J.~X. Tang.
\newblock Viscoelastic properties of semiflexible filamentous bacteriophage fd.
\newblock {\em Phys. Rev. E}, 62(4):5509--5517, 2000.

\bibitem{Tang95}
J.~Tang and S.~Fraden.
\newblock Isotropic-cholesteric phase transition in colloidal dispersions of
  filamentous bacteriophage \textit{fd}.
\newblock {\em Liquid Crystals}, 19(4):459--467, 1995.

\bibitem{taratuta85}
V.~G. Taratuta, A.~J. Hurd, and R.~B. Meyer.
\newblock Light-scattering study of a polymer nematic liquid crystal.
\newblock {\em Phys. Rev. Lett.}, 55(2):246--249, 1985.

\bibitem{Vermant94}
J.~Vermant, P.~Moldenaers, J.~Mewis, and S.~J. Picken.
\newblock Band formation upon cessation of flow in liquid-crystalline polymers.
\newblock {\em J. Rheol.}, 38(5):1571--1589, 1994.

\bibitem{Vermant94b}
J.~Vermant, P.~Moldenaers, S.~J. Picken, and J.~Mewis.
\newblock A comparison between texture and rheological behaviour of lyotropic
  liquid crystalline polymers during flow.
\newblock {\em J. Non-Newt. Fl. Mech.}, 53:1--23, 1994.

\bibitem{Walker96b}
L.~M. Walker, M.~Mortier, and P.~Moldenaers.
\newblock Concentration effects on the rheology and texture of pbg/m-cresol
  solutions.
\newblock {\em J. Rheol.}, 40(5):967--981, 1996.

\bibitem{Walker94}
L.~M. Walker and N.~J. Wagner.
\newblock Rheology of region i flow in a lyotropic liquid-crystal polymer: The
  effects of defect texture) under shear and during relaxation.
\newblock {\em J. Rheol.}, 38(5):1524--1547, 1994.

\end{thebibliography}
\end{document}